\documentclass[%
 aip, 
 amsmath,amssymb,
 reprint,%
]{revtex4-1}

\usepackage{graphicx}
\usepackage{dcolumn}
\usepackage{bm}

\usepackage[utf8]{inputenc}
\usepackage[T1]{fontenc}
\usepackage{mathptmx}
\usepackage{etoolbox}
\usepackage{hyperref}
\usepackage[normalem]{ulem}
\usepackage{dcolumn}
\usepackage{multirow} 
\usepackage{xcolor}
\usepackage{autobreak}

\makeatletter
\def\@email#1#2{%
 \endgroup
 \patchcmd{\titleblock@produce}
  {\frontmatter@RRAPformat}
  {\frontmatter@RRAPformat{\produce@RRAP{*#1\href{mailto:#2}{#2}}}\frontmatter@RRAPformat}
  {}{}
}%
\makeatother
\begin{document}

\preprint{AIP/123-QED}

\title{Does the full configuration interaction method based on quantum phase estimation with Trotter decomposition satisfy the size consistency condition?}
\author{Kenji Sugisaki}
\email{ksugisaki@keio.jp}
\affiliation{Graduate School of Science and Technology, Keio University, 7-1 Shinkawasaki, Saiwai-ku, Kawasaki, Kanagawa 212-0032, Japan}
\affiliation{Quantum Computing Center, Keio University, 3-14-1 Hiyoshi, Kohoku-ku, Yokohama, Kanagawa 223-8522, Japan}
\affiliation{Centre for Quantum Engineering, Research and Education, TCG Centres for Research and Education in Science and Technology, Sector V, Salt Lake, Kolkata 700091, India}

\date{\today}

\begin{abstract}
Electronic structure calculations of atoms and molecules are considered to be a promising application for quantum computers. Two key algorithms, the quantum phase estimation (QPE) and the variational quantum eigensolver (VQE), have been extensively studied. The condition that the energy of a dimer consisting of two monomers separated by a large distance should be equal to twice the energy of a monomer, known as size consistency, is essential in quantum chemical calculations. Recently, we reported that the size consistency condition can be violated by Trotterization in the unitary coupled cluster singles and doubles (UCCSD) ansatz in VQE when employing molecular orbitals delocalized to the dimer (K. Sugisaki {\it et al.}, {\it J. Comput. Chem.}, published online; \href{https://doi.org/10.1002/jcc.27438}{DOI:10.1002/jcc.27438}). It is well known that the full configuration interaction (full-CI) energy is invariant to arbitrary rotations of molecular orbitals, and therefore the QPE-based full-CI should theoretically satisfy the size consistency. However, Trotterization of the time evolution operator can break the size consistency conditions. In this work, we investigated whether size consistency can be maintained with Trotterization of the time evolution operator in QPE-based full-CI calculations. Our numerical simulations revealed that size consistency in QPE-based full-CI is not automatically violated by using molecular orbitals delocalized to the dimer, but employing an appropriate Trotter decomposition condition is crucial to maintain size consistency. We also report on the acceleration of QPE simulations through the sequential addition of ancillary qubits.
\end{abstract}

\maketitle

\section{Introduction} 
Solving the Schr{\"o}dinger equation for atoms and molecules as accurately as possible is a fundamental goal of quantum chemistry. The full configuration interaction (full-CI) method provides the best variational wave functions and energies within the Hilbert space spanned by the chosen basis set. However, the computational cost of the full-CI method grows exponentially with the number of basis functions and electrons, making it impractical except for small molecules with simple basis sets. The method of solving the full-CI on a quantum computer using the quantum phase estimation (QPE) algorithm was proposed in 2005.\cite{AAG-2005} Since then, a number of theoretical studies on QPE-based quantum chemical calculations\cite{Veis-2010, Reiher-2017, Babbush-2018, Lee-2021, Bauman-2021, Kim-2022, Izsak-2022, Kang-2022, Hadler-2022, Casares-2022, Ino-2023, Otten-2024} and extensions of the QPE algorithm\cite{Roggero-2019, Sugisaki-2021, Sugisaki-2023, Kowalski-2024, Sakuma-2024} have been reported. Experimentally, iterative QPE-based full-CI/STO-3G calculations of the H$_2$ molecule were reported in 2010 with photonic\cite{Lanyon-2010} and NMR\cite{Du-2010} quantum processors, using one qubit for wave function encoding. The same system was investigated with two qubits for wave function storage with superconducting qubits in 2016\cite{OMalley-2016}, and on an ion-trap quantum processor in conjunction with a Bayesian QPE framework with quantum error detection.\cite{Yamamoto-2024} The experimental demonstration of the full-CI/STO-3G of the HeH$^+$ molecule was also reported using the NV center of the diamond system.\cite{Wang-2015} Recently, statistical QPE with up to six-qubit Hamiltonian on a superconducting quantum processor was reported.\cite{Blunt-2023} 

In 2014, the quantum--classical hybrid approach known as a variational quantum eigensolver (VQE) was proposed for quantum chemical calculations using noisy intermediate-scale quantum devices.\cite{Peruzzo-2014, Tilly-2022} In VQE, an approximate wave function is generated on a quantum computer using a parameterized quantum circuit defined by an ``ansatz'', and the energy expectation value is computed by statistically sampling the measurement results of the quantum circuit The unitary coupled cluster (UCC) ansatz\cite{Anand-2022} is often used as a physically motivated ansatz, and is defined as
\begin{eqnarray}
    |\Psi_{\mathrm{UCC}}\rangle = e^{(T - T^\dagger)}|\Psi_{\mathrm{HF}}\rangle,
    \label{eq:eq1}
\end{eqnarray}
where $|\Psi_{\mathrm{HF}}\rangle$ is the Hartree--Fock (HF) wave function, and $T$ and $T^\dagger$ are the excitation and de-excitation operators, respectively, applied to the HF wave function. The UCC ansatz, when considering single and double excitation operators as $T$, is known as the UCCSD ansatz and typically provides accurate correlation energies for closed shell molecules. 

In quantum chemical calculations, the energy of a dimer consisting of two monomers separated by a large distance should be twice the energy of a monomer. This condition, known as size consistency,\cite{Szabo-Ostlund} is crucial, especially in the calculation of large molecules. Truncated configuration interaction expansions, such as CISD and CISDT, do not satisfy the size consistency, while truncated coupled cluster methods like CCSD and CCSDT do. The UCCSD ansatz, used in the VQE framework, is based on the cluster expansion and it generally assumed to be size consistent. However, in VQE, the UCCSD ansatz is usually implemented using the Trotter decomposition to construct the parameterized quantum circuit. Recently, we have numerically demonstrated that the Trotterized UCCSD ansatz does not automatically satisfy the size consistency condition.\cite{Sugisaki-2024} Size consistency of the UCCSD ansatz can be maintained when molecular orbitals localized to each monomer are used, but it can be broken when molecular orbitals delocalized over the dimer are employed in the UCCSD ansatz. 

Since QPE-based full-CI is often implemented using the Trotter decomposition of the time evolution operator, it is important to verify whether QPE-based full-CI with Trotter decomposition automatically satisfies the size consistency condition. While it is known that full-CI energy is invariant with respect to the choice of reference molecular orbitals, it is unclear how the robustness of QPE-based full-CI energy is affected by Trotter errors based on the choice of molecular orbitals. It is important to note that Trotter decomposition impacts size consistency differently in VQE-UCCSD and QPE. In VQE-UCCSD, the total energy is calculated as the expectation value of the Hamiltonian, and the contamination of wave function from other electronic states due to Trotter decomposition affects size consistency. In contrast, QPE calculates the total energy based on the phase shift induced by time evolution. In this case, changes in the amount of phase shift due to Trotter error can lead to a breakdown in size consistency.

In this work, we performed numerical quantum circuit simulations of QPE-based full-CI using both localized and delocalized molecular orbitals with various Trotter decomposition conditions, focusing on the size consistency condition. The paper is organized as follows: Section II introduces the QPE-based full-CI method. Section III describes the target molecular systems in this work and outlines the numerical simulation conditions. We also discuss accelerating the numerical simulation of the QPE quantum circuit based on the sequential addition of ancillary qubits. Section IV presents the results of the numerical simulations. A summary of this work is given in Section V.  

\section{Theory}
The QPE is a quantum algorithm designed to find the eigenvalues and corresponding eigenvectors of a unitary matrix $U$ in polynomial time.\cite{Kitaev-1995, Abrams-1999} In quantum chemical calculations, the QPE is performed using the time evolution operator as the unitary operator, 
\begin{eqnarray}
    U = e^{-iHt}. 
    \label{eq:eq2}
\end{eqnarray}
The Born--Oppenheimer approximation\cite{Born-Oppenheimer} is usually adopted, and the electronic Hamiltonian is used as $H$. In the second quantized formula, the electronic Hamiltonian $H$ is expressed as 
\begin{eqnarray}
    H = \sum_{pq}h_{pq} a_p^\dagger a_q + \frac{1}{2}\sum_{pqrs}g_{pqrs} a_p^\dagger a_q^\dagger a_s a_r.
    \label{eq:eq3}
\end{eqnarray}
Here, $h_{pq}$ and $g_{pqrs}$ are the one- and two-electron integrals, respectively, and the indices $p, q, r$, and $s$ run over spin orbitals in the active space. 

\begin{figure}
\includegraphics[width=85mm]{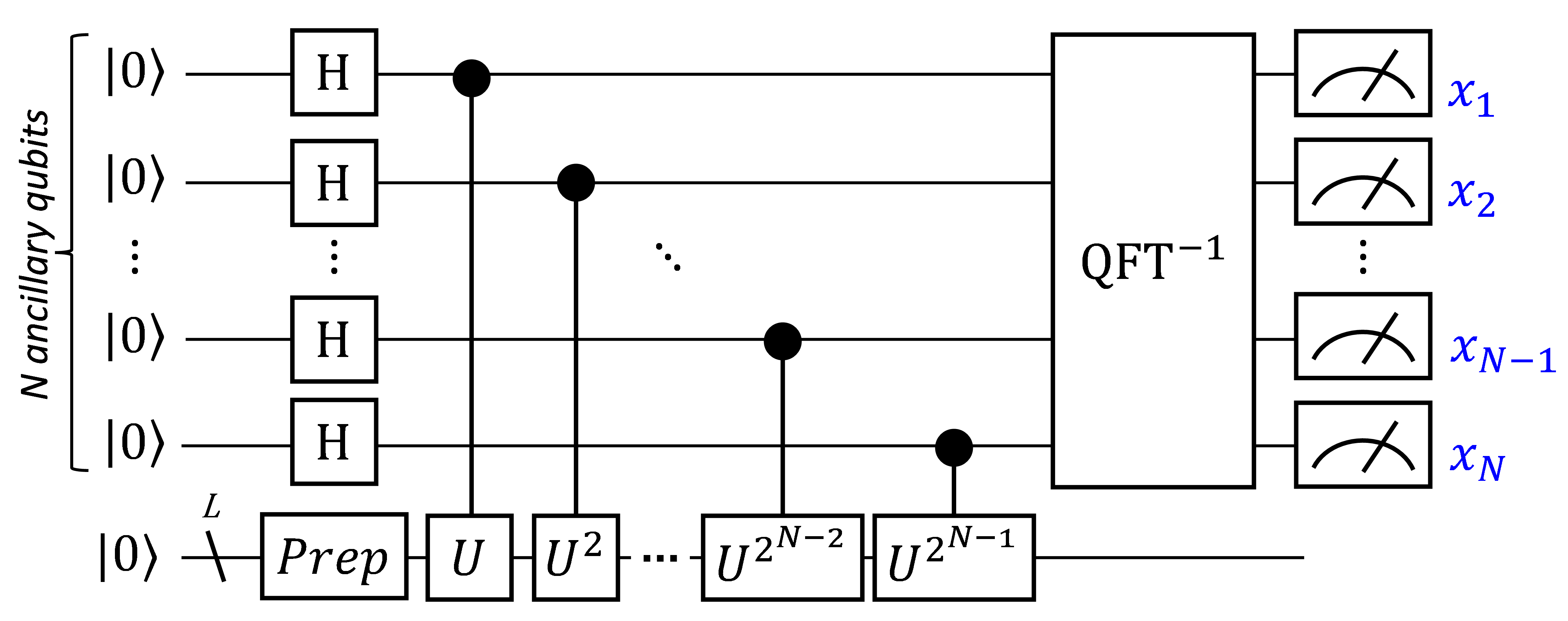}
\caption{\label{fig:fig1} The quantum circuit of the textbook implementation of QPE. $x_1, x_2, \cdots, x_N$ represent the measurement outcomes of ancillary qubits.}
\end{figure}

The textbook implementation of the quantum circuit for QPE\cite{Nielsen-Chuang} is shown in FIG.~\ref{fig:fig1}. This quantum circuit contains $L$ qubits for wave function storage and $N$ ancillary qubits for eigenphase readout.  
The QPE-based full-CI consists of five steps: (1) preparation of the approximate wave function of the target electronic state using a $Prep$ circuit, (2) generation of quantum superposition states using Hadamard (H) gates, (3) controlled-time evolution operation, (4) inverse quantum Fourier transform, and (5) measurement of ancillary qubits. The bit string obtained from the measurement in step (5) corresponds to the fractional binary representation of the eigenphase $\phi = 0.x_1 x_2 \cdots x_N$ in 
\begin{eqnarray}
    e^{-iHt}|\Psi\rangle = e^{-iEt}|\Psi\rangle = e^{i2\pi\phi}|\Psi\rangle, 
    \label{eq:eq4}
\end{eqnarray}
which corresponds to one of the eigenstates. The probability of obtaining a particular eigenvalue is proportional to the overlap between the input and full-CI wave functions. 

In the quantum circuit implementation, the second quantized Hamiltonian in eq. (\ref{eq:eq3}) is transformed into a qubit Hamiltonian $H_q$, 
\begin{eqnarray}
    H_q = \sum_{j=1}^J w_j P_j, 
    \label{eq:eq5}
\end{eqnarray}
using the fermion--qubit transformation techniques such as the Jordan--Wigner transformation (JWT)\cite{JWT} and the Bravyi--Kitaev transformation (BKT)\cite{Seeley-2012}. Here, $P_j$ represents a tensor product of Pauli operators, known as Pauli strings, $J$ is the number of Pauli strings, and $w_j$ is the coefficient computed from $h_{pq}$ and $g_{pqrs}$. The time evolution operator in eq. (\ref{eq:eq2}) is then implemented using the Trotter decomposition. The first-order and second-order Trotter decompositions are given by
\begin{eqnarray}
    e^{-iH_q t} = \left[\Pi_{j=1}^J e^{-i w_j P_j t/M}\right]^M  
    \label{eq:eq6}
\end{eqnarray}
and 
\begin{eqnarray}
    e^{-iH_q t} = \left[\left(\Pi_{j=1}^J e^{-i w_j P_j t/{2M}}\right)\left(\Pi_{j=J}^1 e^{-i w_j P_j t/{2M}}\right)\right]^M, 
    \label{eq:eq7}
\end{eqnarray}
respectively. Here, $M$ is the number of Trotter slices. Besides the Trotter decomposition-based implementation, approaches based on the truncated Taylor series\cite{Berry-2015, Daskin-2018} and qubitization\cite{Qubitization} have also been proposed. The technique based on qubitization shows better computational cost scaling compared to Trotterization, but it requires additional ancillary qubits.\cite{Berry-2019} In this work, we focus on the Trotter decomposition-based approach.

It is well known that the full-CI energy is invariant under arbitrary unitary rotations of the molecular orbitals. For convenience, we focus on the HF canonical molecular orbitals (CMO) and the localized molecular orbitals (LMO) as the two reference orbitals. By defining a unitary matrix $V$ that transforms the two different orbitals, the Hamiltonian matrices in the LMO and CMO bases are written as 
\begin{eqnarray}
    H_\mathrm{LMO} = V^\dagger H_\mathrm{CMO} V. 
    \label{eq:eq8}
\end{eqnarray}
Similarly, the time evolution operators can be expressed as 
\begin{eqnarray}
    U_\mathrm{LMO} = V^\dagger U_\mathrm{CMO} V.
    \label{eq:eq9}
\end{eqnarray}
However, in the full-CI calculations using QPE with Trotter decomposition, the Trotterized time evolution operator $U^{\mathrm{Trotter}}$ does not necessarily satisfy the equality in eq. (\ref{eq:eq9});
\begin{eqnarray}
    U_\mathrm{LMO}^\mathrm{Trotter} \neq V^\dagger U_\mathrm{CMO}^\mathrm{Trotter} V, 
    \label{eq:eq10}
\end{eqnarray}
because the order of the Pauli strings in the Trotterized time evolution operator can depend on the molecular orbitals. We expect that the Pauli strings must be ordered appropriately so that the Trotter error affects both the monomer and the dimer in the same way to satisfy the size consistency condition. 

\begin{figure}
\includegraphics[width=85mm]{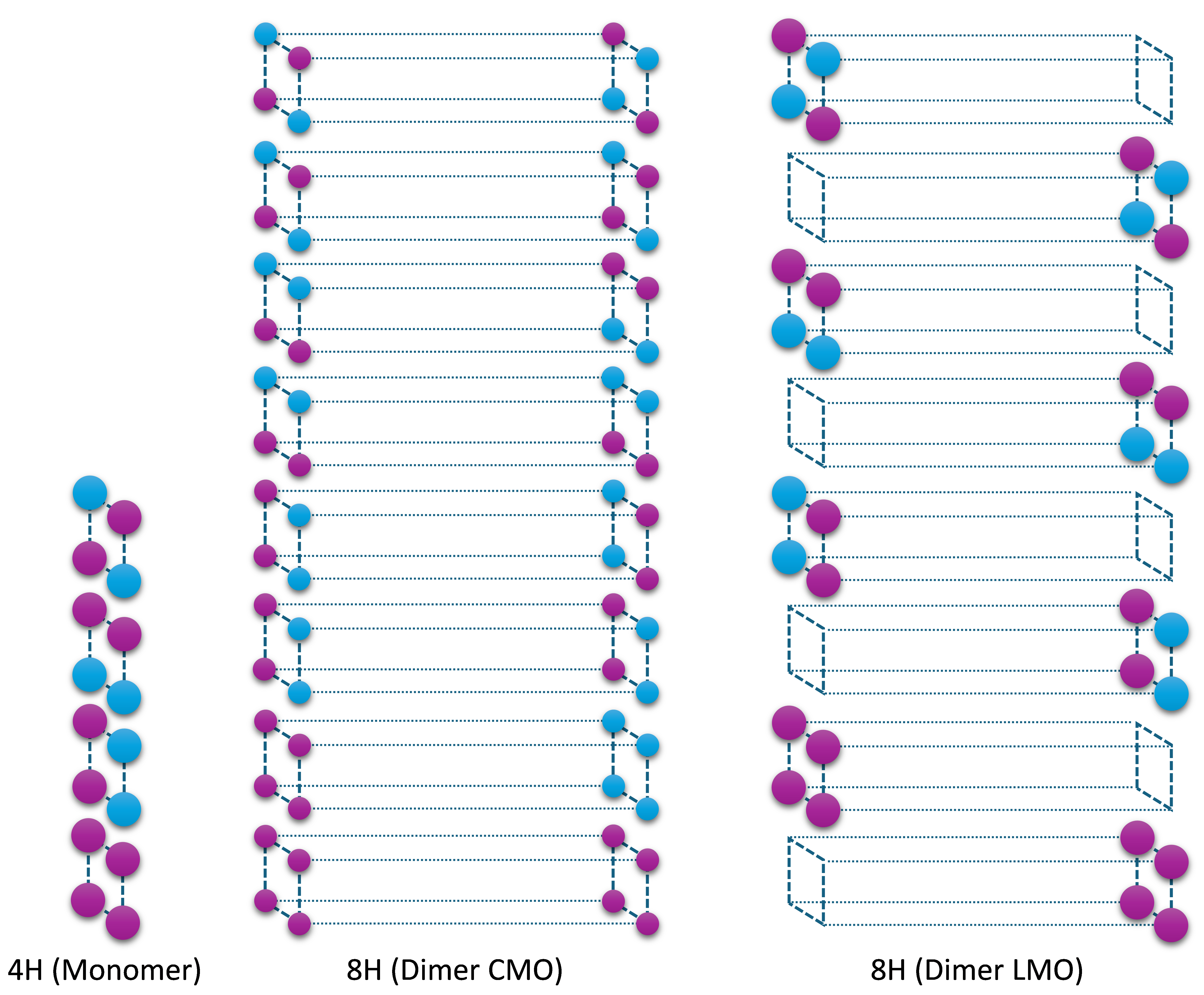}
\caption{\label{fig:fig2} Schematic illustrations of the molecular orbitals for the 4H and 8H clusters. The intermonomer spacing is reduced for clarity. The lowest two and four orbitals for the monomer and the dimers, respectively, are doubly occupied in the HF wave function.}
\end{figure}

\section{Computational conditions}
In this work we mainly focus on a tetrahydrogen (4H) cluster in a square geometry with R(H--H) = 1.0583 \AA~(2.0 Bohr) as the monomer, which is well known as a strongly correlated system.\cite{Paldus-1993} The octahydrogen (8H) cluster, representing the dimer, is constructed by placing two 4H clusters to form a cuboid, with an intermonomer distance of 100 \AA. This system is also used in our previous work.\cite{Sugisaki-2024} We employed the STO-3G basis set, and the active space includes all the molecular orbitals. The size of the active space is (4e, 4o) and (8e, 8o) for 4H and 8H clusters, respectively. Here, ($k$e, $l$o) indicates that the active space contains $k$ electrons and $l$ molecular orbitals. 

In the present study, we used two different sets of molecular orbitals for the dimer calculations: the CMOs, delocalized to the dimer under the D$_{\mathrm{2h}}$ point group, and the LMOs under the C$_{\mathrm{2v}}$ point group. Schematic illustrations of the CMO and LMO of 4H and 8H clusters are given in FIG.~\ref{fig:fig2}. These molecular orbitals are generated using the GAMESS-US software~\cite{GAMESS}.  

Conventional fermion--qubit transformations such as JWT and BKT require 2$l$ qubits to encode the wave function. In this work, we used the symmetry-conserving Bravyi--Kitaev transformation (SCBKT),\cite{Bravyi-2017} which reduces the qubit count by two by specifying the number of spin-$\alpha$ and spin-$\beta$ electrons. Thus, the number of qubits for wave function encoding ($L$ in FIG.~\ref{fig:fig1}) is 6 and 14 for 4H and 8H clusters, respectively. 

{In addition to the 4H/8H clusters, we also studied 2H/4H clusters to examine the system size dependence, and the acetylene (HC$\equiv$CH) molecule under triple bond dissociation as the representative system of covalent bond cleavage. For the 2H/4H clusters calculations, the intra- and inter-molecular H--H distances are set to 1.0583 \AA~and 100 \AA, respectively. The STO-3G basis set is used, with the active space being (2e, 2o) and (4e, 4o) for the monomer and dimer, respectively. In the case of acetylene, the monomer is the lowest spin-quartet state of the C--H fragment, with a bond length set to the experimental value of acetylene (1.1199 \AA). Using the STO-3G basis set and freezing the 1s core orbital of the C atoms, we constructed (5e, 5o) and (10e, 10o) active spaces for the monomer and dimer, respectively.}

In the QPE simulations, we set the time length of the time evolution operator in eq. (\ref{eq:eq2}) to $t = 1.0$. We used 10 ancillary qubits to read out the eigenphase for 4H/8H and 2H/4H clusters, and 8 ancillary qubits for the triple bond dissociation of acetylene. As input wave functions for the QPE, we examined both the full-CI and HF wave functions to assess the state dependence of the Trotter error. For the implementation of the controlled-time evolution operator, we utilized both the first-order and second-order Trotter decompositions with $M$ = 1, 2, 5, and 10. The Trotter error depends on the order of the Pauli strings. Unless otherwise stated, we employed magnitude ordering, where the Pauli strings are applied in descending order of the absolute value of the corresponding coefficient, $|w_j|$.\cite{Tranter-2018} All simulations were conducted using our custom Python3 code, developed with the OpenFermion,\cite{OpenFermion} Cirq,\cite{Cirq} and cuQuantum\cite{cuQuantum} libraries. For reference, the Trotter-free time evolution was implemented using the \texttt{expm} function in the SciPy library.\cite{SciPy} 

\begin{figure*}[t]
    \includegraphics[width=140mm]{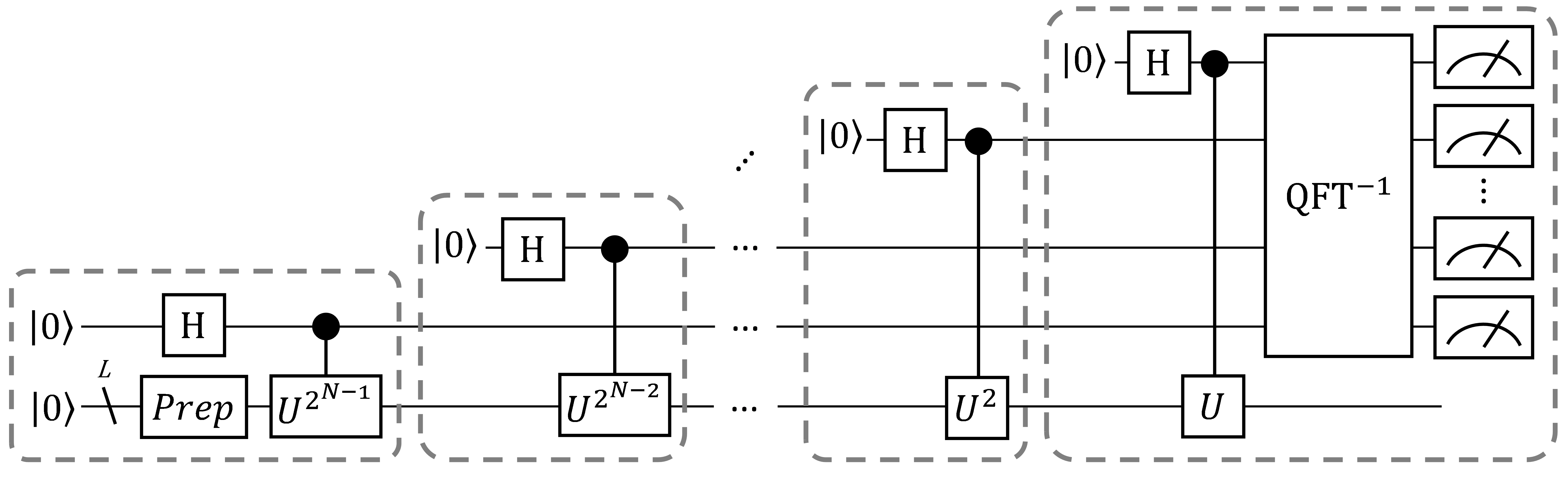}
    \caption{Schematic illustration of the numerical simulations for  the QPE-based full-CI with four ancillary qubits. Each step of the numerical simulation is indicated by a dotted square.}
    \label{fig:fig3}
\end{figure*}

The computational cost of quantum circuit simulations for QPE-based full-CI is substantial, making acceleration crucial for studying larger systems. In QPE with $N$ ancillary qubits, we need to perform the controlled-$U$ operation $2^N - 1$ times, which is the most time-consuming process. In this study, we reduce the simulation cost by sequentially adding ancillary qubits one by one. The schematic representation of our QPE quantum circuit simulation is shown in FIG.~\ref{fig:fig3}. In this strategy, the most time-consuming controlled-$U^{2^{N-1}}$ is simulated with $L + 1$ qubits in the first step (the leftmost dotted square in FIG.~\ref{fig:fig3}). Next, we add an ancillary qubit to simulate the controlled-$U^{2^{N-2}}$ with $L + 2$ qubits in the second step (shown in the second dotted square from the left in FIG.~\ref{fig:fig3}). In this approach, the simulation time for the second step is approximately the same as for the first step. This occurs because the depth of the quantum circuit in the second step is about half of that in the first step, while the number of qubits increases by one. The reduction in simulation time due to the decreased circuit depth is offset by the increased qubit count, leading to nearly equal simulation times for both steps.The QPE simulations based on the implementation in FIG.~\ref{fig:fig3}, resulting in an exponential speed up of the numerical simulations from naive implementation in FIG.~\ref{fig:fig1}.  

\begin{figure}
\includegraphics[width=85mm]{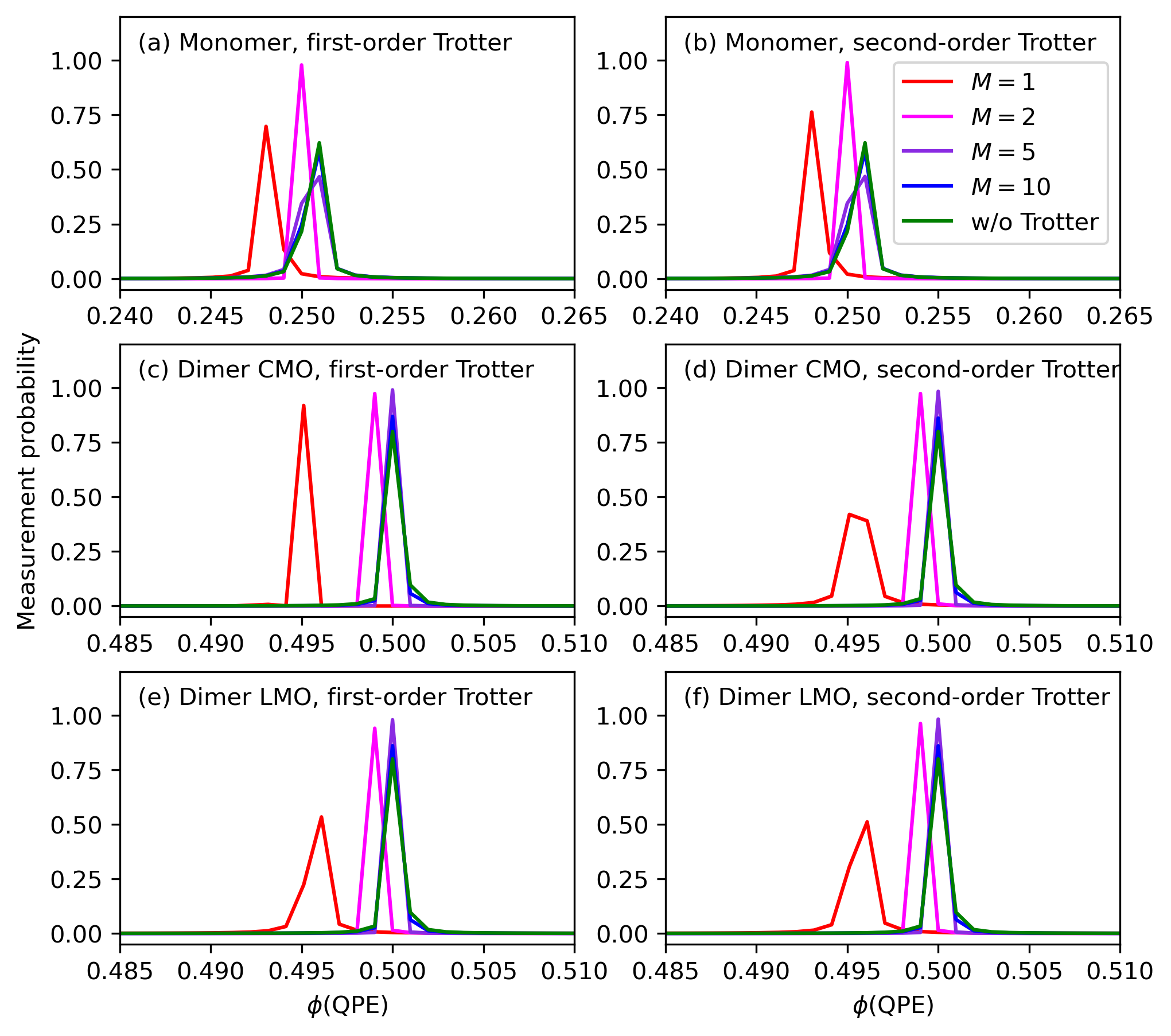}
\caption{\label{fig:fig4} Plot of phase value versus measurement probability from the QPE simulation of 4H/8H clusters using the full-CI wave function as the input. Magnitude ordering is used for the Trotter decomposition.}
\end{figure}

\begin{figure}
\includegraphics[width=85mm]{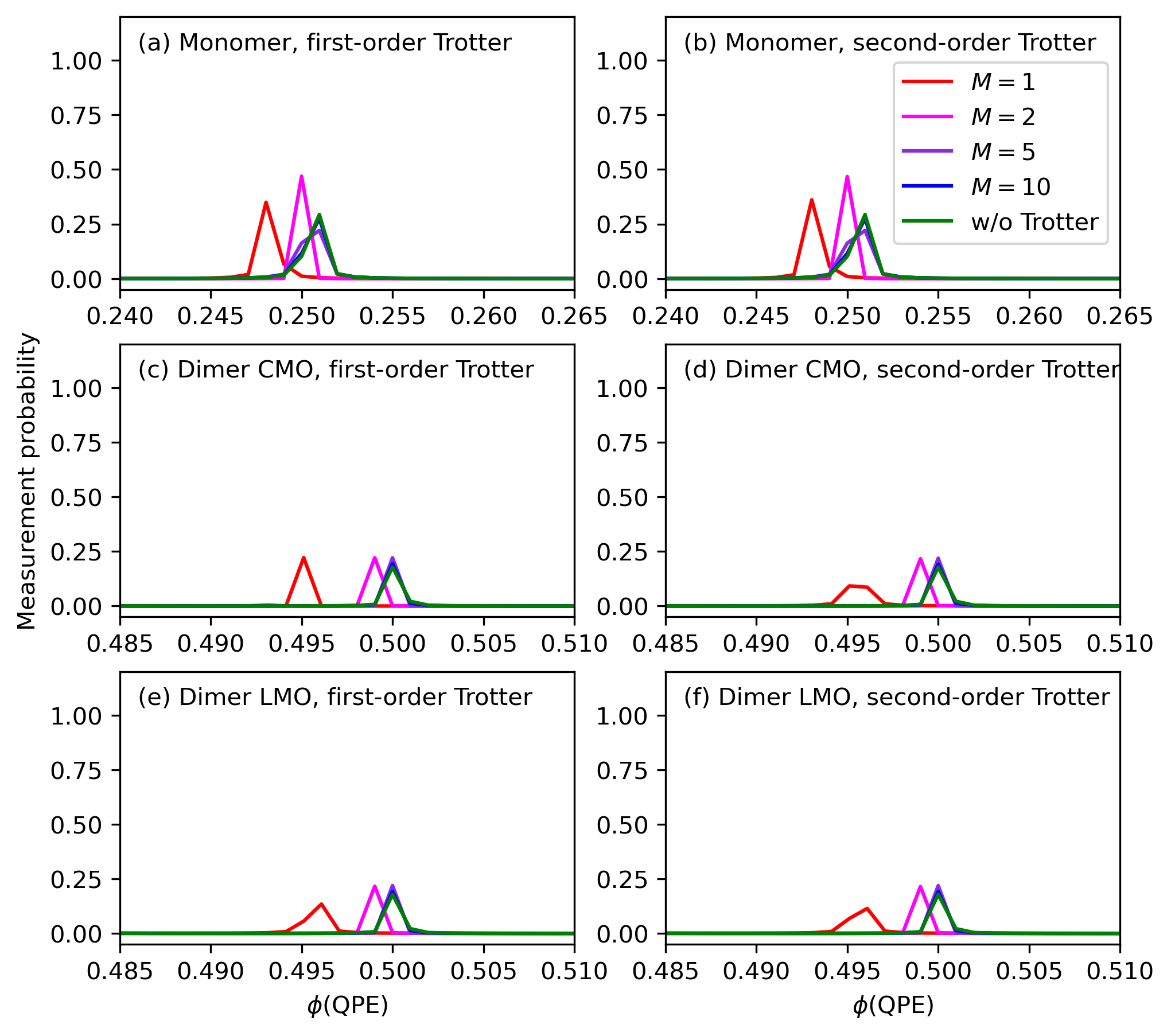}
\caption{\label{fig:fig5} Plot of phase value versus measurement probability from the QPE simulation of 4H/8H clusters using the HF wave function as the input. Magnitude ordering is used for the Trotter decomposition.}
\end{figure}

\section{Results}
\subsection{4H and 8H clusters}
Plots of phase value versus measurement probability from the QPE-based full-CI of the 4H and 8H clusters, as a function of the number of Trotter slices $M$, are shown in FIG.~\ref{fig:fig4} and \ref{fig:fig5}, using full-CI and HF wave functions, respectively, as the inputs. For small $M$, the peak obtained from the Trotterized time evolution operator is shifted from the peak calculated by the Trotter-free simulations, but it converges to the Trotter-free result as $M$ increases. Notably, both CMO- and LMO-based QPE simulations yield peaks at nearly the same position. The peak height is reduced when using the HF wave function as the input due to the decreased overlap between the input and full-CI wave functions, although the peak position remains consistent between the full-CI and HF inputs. It is important to note that some peaks exhibit larger variance due to the well-known ``leakage problem'' in the textbook implementation of QPE.\cite{Xiong-2022} 

\begin{figure}
\includegraphics[width=85mm]{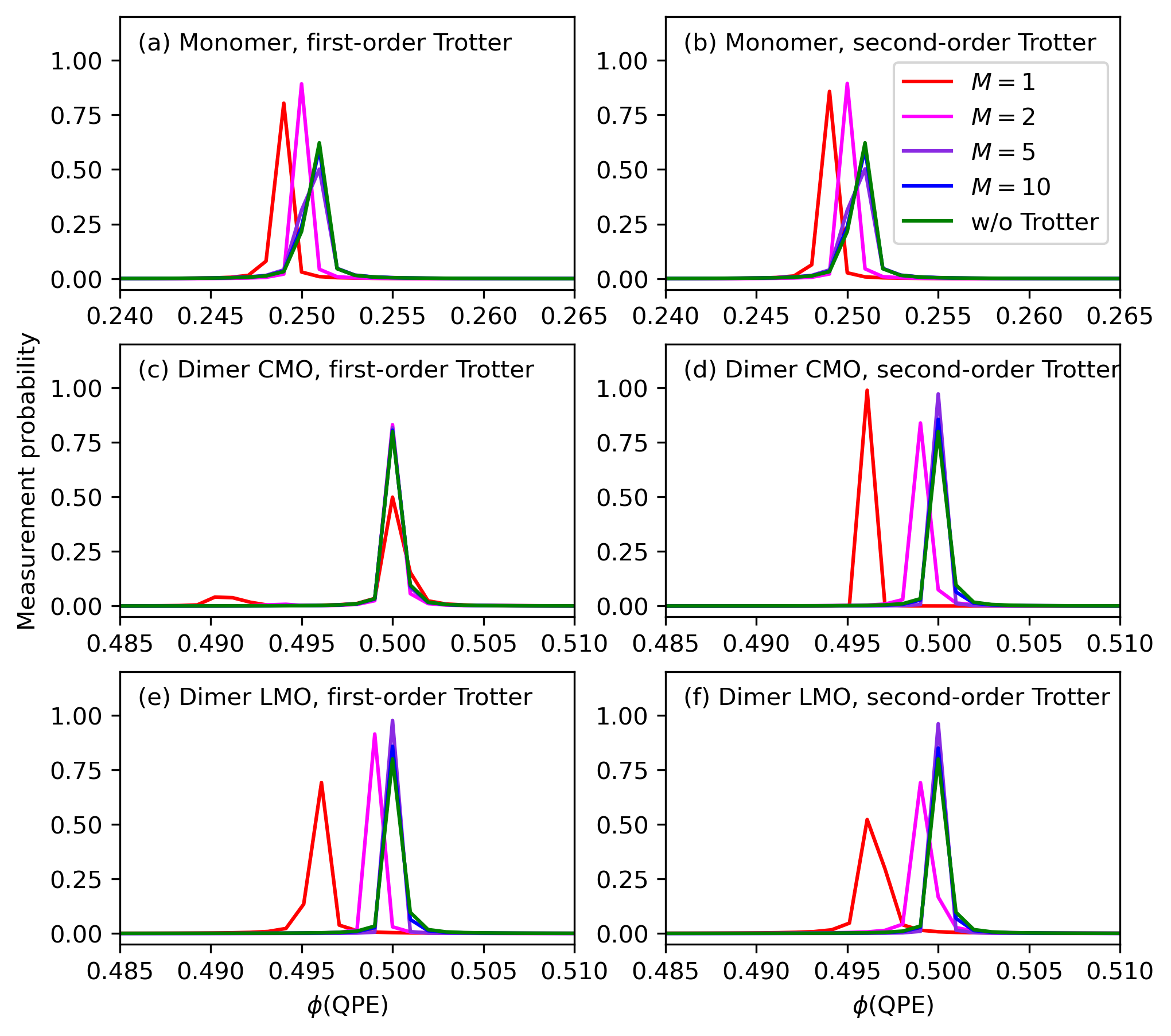}
\caption{\label{fig:fig6} Plot of phase value versus measurement probability from the QPE simulation of 4H/8H clusters with the full-CI wave function as the input. Lexicographic ordering is used for the Trotter decomposition.}
\end{figure}

To assess the impact of operator ordering on the Trotterized time evolution operator, we examined lexicographic ordering in the Trotter decomposition. The results are shown in FIG.~\ref{fig:fig6}. As anticipated, the trend of the Trotter error varied with different strategies for operator ordering. In the dimer CMO calculations using the first-order Trotter decomposition, the computed peak is already quite close to the Trotter-free value even for $M$ = 1, likely due to a fortuitous cancellation of errors (see FIG.~\ref{fig:fig6}(c)). However, in the first-order Trotter simulations for $M$ = 1 in the dimer CMO, an additional small peak is observed around $\phi$ = 0.492, despite using the full-CI wave function as the input. This is because the Trotterized time evolution operator differs from the original time evolution operator, and the full-CI wave function is no longer an eigenfunction of the Trotterized time evolution operator. The presence of additional peaks indicates a significant deviation of $U^\mathrm{Trotter}$ from $U$, suggesting that a more refined approach to Trotter decomposition is necessary. It should be noted that the peak positions are insensitive to the order of the Trotter decomposition for monomer and dimer with LMO. However, this is not always the case. In fact, as we discuss in the following subsections, we observed clear differences in size consistency breakdown between first-order and second-order Trotter decompositions. 

\begin{figure}
\includegraphics[width=85mm]{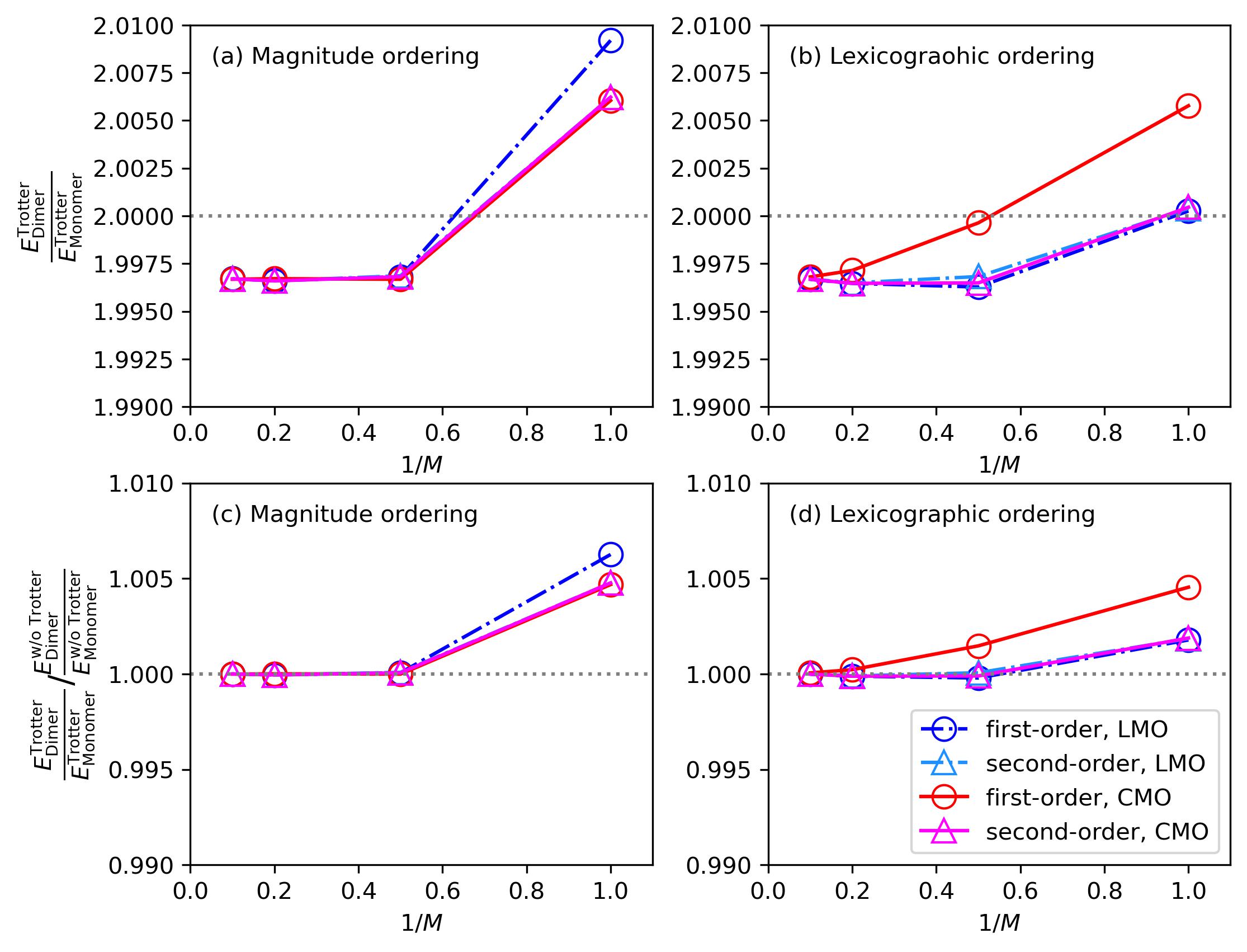}
\caption{\label{fig:fig7} Ratio of the total energies of the 8H cluster (dimer) to the 4H cluster (monomer) calculated using (a) magnitude ordering and (b) lexicographic ordering. The ratio values normalized by the Trotter-free simulation results are plotted in (c) and (d). }
\end{figure}

The eigenphase value is determined by fitting the measurement probability plot with a Gaussian function. The ratio of the dimer and monomer eigenenergies, $E_\mathrm{Dimer}/E_\mathrm{Monomer}$, is computed and plotted in FIG.~\ref{fig:fig7}(a) and (b) for magnitude and lexicographic orderings, respectively, as a function of $1/M$. Ideally, when size consistency is perfectly maintained, this ratio should be equal to two. However, in our numerical simulations, the ratio deviates slightly from two even in the Trotter-free implementation, likely due to rounding and leakage errors. To assess the deviation of $E_\mathrm{Dimer}/E_\mathrm{Monomer}$ values from Trotter decomposition-based simulation (denoted as ``Trotter'' in the superscript) compare to Trotter-free simulations (denoted as ``w/o Trotter'' in the superscript), we plot $E_\mathrm{Dimer}^\mathrm{Trotter}/E_\mathrm{Monomer}^\mathrm{Trotter}$ normalized by $E_\mathrm{Dimer}^\mathrm{w/o\ Trotter}/E_\mathrm{Monomer}^\mathrm{w/o\ Trotter}$ in FIG.~\ref{fig:fig7}(c) and (d) for magnitude and lexicographic orderings, respectively. In this case, using $M$ = 1 for the Trotter decomposition (where $\Delta t$, the time length of a single Trotter slice, is 1.0) is insufficient to meet the size consistency condition, irrespective of the operator ordering method and the molecular orbitals employed for the wave function expansion. Increasing the number of Trotter slices to $M$ = 2 nearly satisfies the size consistency condition, except for the CMO with first-order Trotter decomposition using lexicographic ordering. Even in this case, size consistency condition is systematically recovered with a higher number of Trotter slices. Our numerical simulations indicate that size consistency is not inherently violated using LMOs for wave function expansion. However, the convergence behavior towards Trotter-free results with respect to the number of Trotter slices underscores the importance of selecting appropriate Trotter decomposition conditions, such as reference molecular orbitals, the order of Trotter decomposition, and operator ordering, to ensure size consistency in QPE-based full-CI calculations. 

\subsection{2H and 4H clusters}
\begin{figure}
\includegraphics[width=85mm]{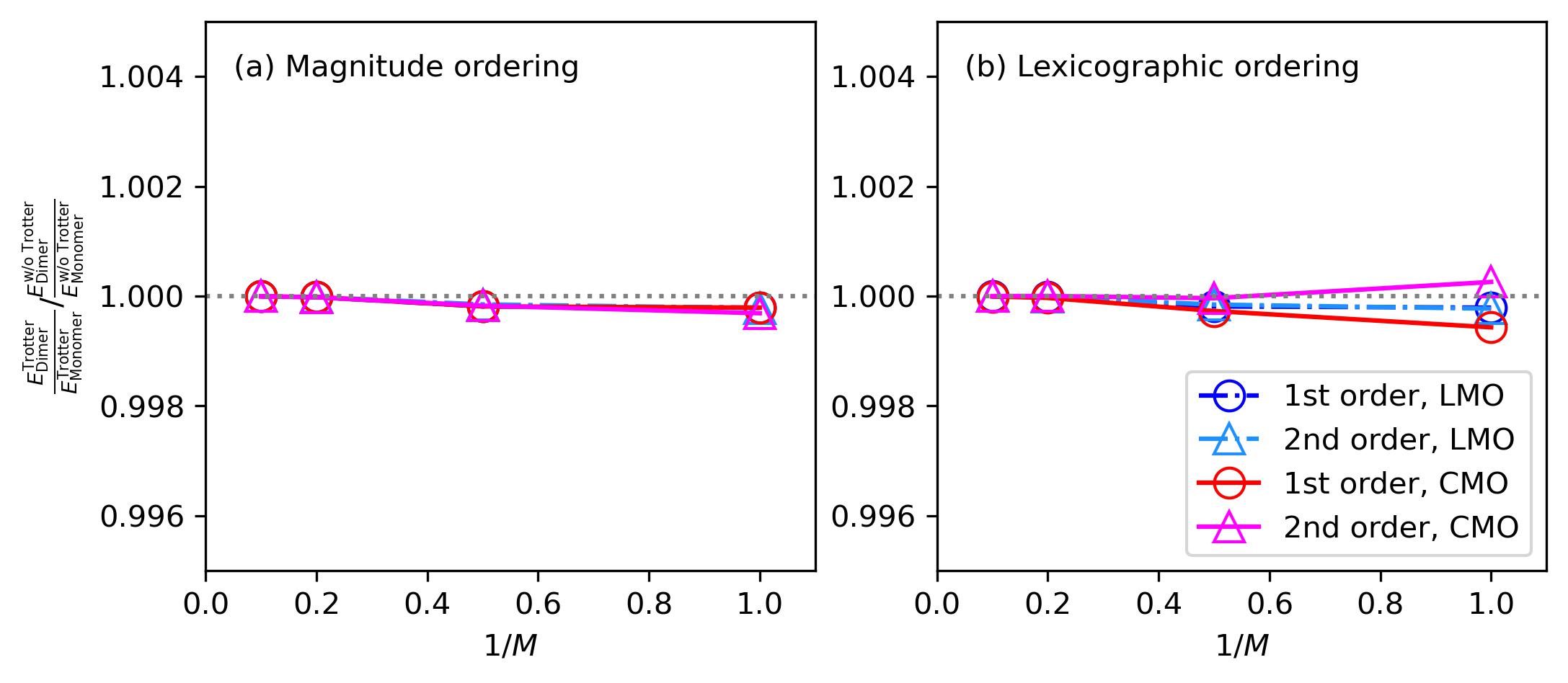}
\caption{\label{fig:fig8} Ratio of the total energies of the 4H cluster (dimer) to the 2H cluster (monomer) normalized by the Trotter-free simulation results. (a) Magnitude ordering. (b) Lexicographic ordering}
\end{figure}

Next, we investigate the system size dependency of the breakdown of size consistency in the QPE with Trotterized time evolution operators. We performed the QPE simulations using 2H and 4H clusters as a monomer and a dimer, respectively. 
The ratio of the total energies of the dimer to the monomer, normalized by the Trotter-free simulation results, is plotted in FIG.~\ref{fig:fig8}. Note that the scale of the vertical axis of FIG.~\ref{fig:fig8} is a half of that of FIG.~\ref{fig:fig7}(c) and (d). Although the deviation from size consistency is smaller in 2H/4H clusters compared to 4H/8H clusters, lexicographic ordering in conjunction with CMO is likely to break size consistency when the number of Trotter slices is small, whereas LMO nearly maintains size consistency with lexicograhic ordering for $M$ = 1. 

\subsection{Triple bond dissociation in acetylene}
\begin{figure}
\includegraphics[width=85mm]{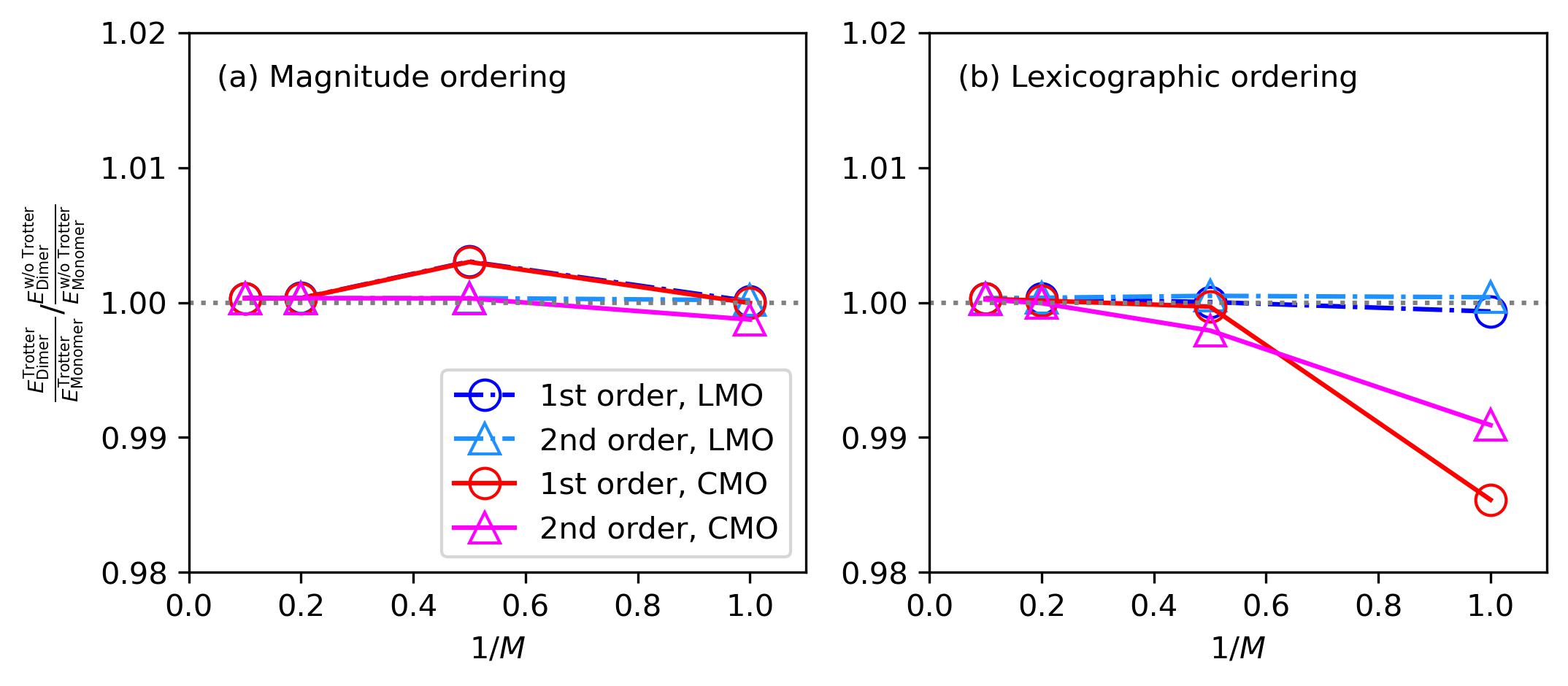}
\caption{\label{fig:fig9} Ratio of the total energies of the dimer to the monomer normalized by the Trotter-free simulation results in triple bond dissociation of acetylene. (a) Magnitude ordering. (b) Lexicographic ordering}
\end{figure}

To examine the size consistency condition under bond breaking, we conducted QPE simulations of acetylene undergoing triple bond dissociation, using the C$\equiv$H fragment as the monomer. The results are summarized in FIG.~\ref{fig:fig9}. Due to the large total energies of the acetylene system, we plotted the ratio of the eigenenergies of the qubit Hamiltonian without constant terms (frozen core energy and the constant term obtained in the SCBKT). Note that the constant term of the dimer is exactly twice that of the monomer. We observed clear difference in the behavior of size consistency breakdown between CMO and LMO. Size consistency is nearly preserved when LMO is used for wave function expansion, even with lexicographic ordering and $M$ = 1 in the Trotter decomposition. In contrast, the CMO-based calculation shows a breakdown in size consistency for small $M$. It is important to note that size consistency does not always guarantee accurate total energy. Specifically, $E_{\rm{Monomer}}^{\rm{w/o Trotter}} - E_{\rm{Monomer}}^{\rm{Trotter}}$ and $E_{\rm{Dimer;LMO}}^{\rm{w/o Trotter}} - E_{\rm{Dimer;LMO}}^{\rm{Trotter}}$ values with lexicographic ordering and $M$ = 1 are calculated as 0.0020 and 0.0040 Hartree, respectively.

\subsection{Comparison between LMO and CMO}
As we observed from the numerical simulations, the extent of size consistency violation is smaller for LMO compared to CMO. This observation is consistent with expectations from a chemical perspective. 

Consider a dimer composed of spatially well-separated monomers A and B. In the LMO basis, in the limit of weak inter-monomer interaction, the Hamiltonian of the dimer, $H(A+B)$, can be expressed as a linear combination of monomer Hamiltonians $H(A)$ and $H(B)$,
\begin{eqnarray}
    H(A+B) = H(A) + H(B)
    \label{eq:eq11}
\end{eqnarray}
with
\begin{eqnarray}
    [H(A), H(B)] = 0.
    \label{eq:eq12}
\end{eqnarray}
Using the commutation relation in eq. (\ref{eq:eq12}), the Trotterized time evolution operator for the dimer can be rewritten as the product of the Trotterized time evolution operators for monomers A and B, with the Trotterized terms arranged accordingly. For instance, in the first-order Trotter decomposition, we have
\begin{align}\label{eq:eq13}
\begin{autobreak}
    e^{-iH(A+B)_q t} 
    = \left[\Pi_j e^{-i w_j P_j t/M}\right]^M 
    = \left[\Pi_{k\in A} e^{-i w_k P_k t/M} \Pi_{l\in B} e^{-i w_l P_l t/M}\right]^M. 
\end{autobreak}
\end{align}
If the orderings of the Pauli strings $P_{k\in A}$ and $P_{l\in B}$ in eq. (\ref{eq:eq13}) are the same as those in the Trotterized time evolution operators for monomers A and B, respectively, then size consistency can be preserved.

In contrast, when CMOs are used for wave function expansion, decomposing the dimer Hamiltonian $H(A+B)$ into $H(A)$ and $H(B)$ is not straightforward. Consequently, it is not possible to rearrange the terms in the Trotterized time evolution operator for the dimer to express it as a product of the Trotterized time evolution operators for the monomers.

\section{Summary}
A recent theoretical study on VQE-UCCSD highlighted that size consistency can be violated by  Trotter decomposition, when the molecular orbitals delocalized over the dimer are used.\cite{Sugisaki-2024} In this study, we investigated the impact of Trotter decomposition of the time evolution operator in QPE-based full-CI calculations, focusing on the size consistency condition. Our numerical simulations of the 4H/8H clusters, 2H/4H clusters, and triple bond dissociation in acetylene, using both CMOs and LMOs, revealed that the eigenphase values obtained from QPE are less sensitive to the choice of molecular orbitals when magnitude ordering is employed, and size consistency is nearly maintained when the evolution time length of a single Trotter slice is set to 0.2 or shorter. In contrast, with lexicographic ordering, size consistency is not satisfied when CMO is used in conjunction with small number of Trotter slices. However, size consistency can be systematically recovered by increasing the number of Trotter slices. The fact that QPE-based full-CI can satisfy the size consistency condition under appropriate Trotter decomposition conditions is promising for quantum chemical calculations on quantum computers. This becomes particularly relevant as large-scale quantum chemical calculations that are intractable on classical computers become feasible. The integration of QPE with fragmentation-based methods where size consistency is crucial, such as divide-and-conquer (DC),\cite{Yang-1995} density matrix embedding theory (DMET),\cite{Knizia-2012} and fragment molecular orbital (FMO)\cite{Kitaura-1999} methods, appears to be a promising direction, which will be discussed in a forthcoming paper. 

\begin{acknowledgments}
The computation was carried out using the JHPCN Joint Research Projects (jh240001) on supercomputer ‘Flow’ at Information Technology Center, Nagoya University. K. Sugisaki acknowledges Naoki Yamamoto, Takashi Abe, Yu-ya Ohnishi, Shu Kanno, and Yudai Suzuki for useful discussions. 
This work was supported by Quantum-LEAP Flagship Program (JPMXS0120319794) from Ministry of Education, Culture, Sports, Science and Technology (MEXT), Japan; Center of Innovations for Sustainable Quantum AI (JPMJPF2221) from Japan Science and Technology Agency (JST), Japan; and Grants-in-Aid for Scientific Research C (21K03407) and for Transformative Research Area B (23H03819) from Japan Society for the Promotion of Science (JSPS), Japan. 
\end{acknowledgments}

\section*{Author Declarations}
\subsection*{Conflict of Interest}
The author has no conflicts to disclose. 

\subsection*{Author Contributions}
{\bf Kenji Sugisaki}: Conceptualization (Lead); Software (Lead); Investigation (Lead); Validation (Lead); Writing -- original draft (Lead); Writing -- review and editing (Lead) 

\section*{Data Availability}
The data that support the findings of this study are available from the corresponding author upon reasonable request.

\section*{References}
\bibliography{refs.bib}

\end{document}